\documentclass[preprint,preprintnumbers,prl,a4paper,superscriptaddress,showpacs]{revtex4}

\usepackage{graphicx}
\usepackage{psfrag}

\newcommand{\GeV}{\,\mbox{GeV}}
\newcommand{\MeV}{\,\mbox{MeV}}

\newcommand{\BelleIntLum}{357}

\newcommand{\BelleNbb}{386}
\newcommand{\BelleNbbErr}{4}
				   
\newcommand{\BFlclckp}{6.5^{+1.0}_{-0.9}\pm1.1\pm3.4}
\newcommand{\BFlclcks}{7.9^{+2.9}_{-2.3}\pm1.2\pm4.1}

\newcommand{\EFFlclckp}{7.79}
\newcommand{\EFFlclcks}{1.38}

\newcommand{\Nexplclckp}{48.5^{+7.5}_{-6.8}}
\newcommand{\Nexplclcks}{10.5^{+3.8}_{-3.1}}

\newcommand{\Sexplclckp}{15.4}
\newcommand{\Sexplclcks}{6.6}

\newcommand{\NexplclckpLcp}{39.5^{+7.3}_{-6.5}}
\newcommand{\NexplclckpLcm}{48.2^{+7.7}_{-7.0}}

\newcommand{\NexplclcksLcp}{11.4^{+3.8}_{-3.2}}
\newcommand{\NexplclcksLcm}{10.0^{+3.8}_{-3.1}}

\newcommand{\ULexplclckpLcsideLcsign}{1.7}
\newcommand{\ULexplclcksLcsideLcsign}{0.2}

\graphicspath{{.}{figures/}}
\begin{document}
  
  \preprint{
\vbox{
      \hbox{Belle preprint 2005-37 } 
      \hbox{KEK preprint   2005-89 }
  }
}

\date{Version 12.2.3}

  \title{ 
    \boldmath 
    Observation of
    ${{B^+}\to\Lambda_c^+{\Lambda}_c^-{K}^+}$ and
    ${{B^0}\to\Lambda_c^+{\Lambda}_c^-{K}^0}$ 
    decays}

\affiliation{Budker Institute of Nuclear Physics, Novosibirsk}
\affiliation{Chonnam National University, Kwangju}
\affiliation{University of Cincinnati, Cincinnati, Ohio 45221}
\affiliation{Deutsches Elektronen--Synchrotron, Hamburg}
\affiliation{University of Hawaii, Honolulu, Hawaii 96822}
\affiliation{High Energy Accelerator Research Organization (KEK), Tsukuba}
\affiliation{Institute of High Energy Physics, Chinese Academy of Sciences, Beijing}
\affiliation{Institute of High Energy Physics, Protvino}
\affiliation{Institute of High Energy Physics, Vienna}
\affiliation{Institute for Theoretical and Experimental Physics, Moscow}
\affiliation{J. Stefan Institute, Ljubljana}
\affiliation{Kanagawa University, Yokohama}
\affiliation{Korea University, Seoul}
\affiliation{Swiss Federal Institute of Technology of Lausanne, EPFL, Lausanne}
\affiliation{University of Ljubljana, Ljubljana}
\affiliation{University of Maribor, Maribor}
\affiliation{University of Melbourne, Victoria}
\affiliation{Nagoya University, Nagoya}
\affiliation{Nara Women's University, Nara}
\affiliation{National Central University, Chung-li}
\affiliation{National United University, Miao Li}
\affiliation{Department of Physics, National Taiwan University, Taipei}
\affiliation{H. Niewodniczanski Institute of Nuclear Physics, Krakow}
\affiliation{Niigata University, Niigata}
\affiliation{Nova Gorica Polytechnic, Nova Gorica}
\affiliation{Osaka City University, Osaka}
\affiliation{Panjab University, Chandigarh}
\affiliation{Peking University, Beijing}
\affiliation{Princeton University, Princeton, New Jersey 08544}
\affiliation{RIKEN BNL Research Center, Upton, New York 11973}
\affiliation{University of Science and Technology of China, Hefei}
\affiliation{Shinshu University, Nagano}
\affiliation{Sungkyunkwan University, Suwon}
\affiliation{University of Sydney, Sydney NSW}
\affiliation{Tata Institute of Fundamental Research, Bombay}
\affiliation{Toho University, Funabashi}
\affiliation{Tohoku Gakuin University, Tagajo}
\affiliation{Tohoku University, Sendai}
\affiliation{Department of Physics, University of Tokyo, Tokyo}
\affiliation{Tokyo Institute of Technology, Tokyo}
\affiliation{Tokyo Metropolitan University, Tokyo}
\affiliation{University of Tsukuba, Tsukuba}
\affiliation{Virginia Polytechnic Institute and State University, Blacksburg, Virginia 24061}
\affiliation{Yonsei University, Seoul}
   \author{N.~Gabyshev}\affiliation{Budker Institute of Nuclear Physics, Novosibirsk} 
   \author{K.~Abe}\affiliation{High Energy Accelerator Research Organization (KEK), Tsukuba} 
   \author{K.~Abe}\affiliation{Tohoku Gakuin University, Tagajo} 
   \author{I.~Adachi}\affiliation{High Energy Accelerator Research Organization (KEK), Tsukuba} 
   \author{H.~Aihara}\affiliation{Department of Physics, University of Tokyo, Tokyo} 
   \author{Y.~Asano}\affiliation{University of Tsukuba, Tsukuba} 
   \author{V.~Aulchenko}\affiliation{Budker Institute of Nuclear Physics, Novosibirsk} 
   \author{T.~Aushev}\affiliation{Institute for Theoretical and Experimental Physics, Moscow} 
   \author{S.~Bahinipati}\affiliation{University of Cincinnati, Cincinnati, Ohio 45221} 
   \author{A.~M.~Bakich}\affiliation{University of Sydney, Sydney NSW} 
% ...
\collaboration{The Belle Collaboration}

\begin{abstract}
  We report first measurements of the doubly charmed baryonic $B$ decays
  ${{B}\to\Lambda_c^+{\Lambda}_c^-{K}}$.
  The ${{B^+}\to\Lambda_c^+{\Lambda}_c^-{K^+}}$ decay is observed with
  a branching fraction of 
  $(\BFlclckp)\times10^{-4}$
  and a statistical significance of $\Sexplclckp\,\sigma$.
  The ${{B^0}\to\Lambda_c^+{\Lambda}_c^-{K}^0}$ decay is observed with
  a branching fraction of $(\BFlclcks)\times10^{-4}$
  and a statistical significance of $\Sexplclcks\,\sigma$.
  The branching fraction errors are statistical, systematic,
  and the error resulting from the uncertainty of the
  $\Lambda_c^+\to{p}K^-\pi^+$ decay branching fraction. 
  The analysis is based on \BelleIntLum\,fb$^{-1}$ of data accumulated
  at the
  $\Upsilon(4S)$ resonance with the Belle detector at the KEKB
  asymmetric-energy $e^+ e^-$ collider.   
\end{abstract}

\pacs{14.20.Lq, 14.40.Nd}

\maketitle

Recently a number of studies of single charmed baryon production in $B$ decays
have been reported~\cite{cleo_blamc,belle_blamc,belle_lamc1,belle_lamc2}.
The measured branching fractions of the two-body single charmed baryon decays
$\bar{B}^0\to\Lambda_c^+\bar{p}$~\cite{belle_lamc1} and 
${B}^-\to\Sigma_c^0(2455)\bar{p}$~\cite{belle_lamc2} are
significantly smaller than theoretical
expectations~\cite{qcd_sum_rule,duquark,pole,bag}. 
The multi-body single charmed baryon decays
$\bar{B}\to\Lambda_c^+\bar{p}\pi(\pi)$ were found to have branching fractions
about one order of magnitude larger than the corresponding two-body
decays, but still below theoretical predictions.
While single charm production proceeds via a $b{\to}c\bar{u}d$
quark transition, production of two charmed particles occurs via
a $b{\to}c\bar{c}s$ transition. 
In contrast to the single charmed baryon production
the two-body doubly charmed baryon $B$ decay 
$B^+\to\bar{\Xi}_c^0\Lambda_c^+$~\cite{belle_xic_lamc} recently
observed at Belle has a branching fraction comparable
to theoretical predictions~\cite{qcd_sum_rule}.
It would be interesting to check whether theory can describe multibody
double charmed decays.
In this paper we report the first observation of the 
${B^{+}}\to\Lambda_c^+{\Lambda}_c^-K^{+}$ and 
${B^{0}}\to\Lambda_c^+{\Lambda}_c^-K^{0}$ decays,
which are three-body decays that proceed via
a $b{\to}c\bar{c}s$ transition.
Inclusion of charge conjugate states is implicit unless otherwise
stated.
The analysis is based on a data sample of \BelleIntLum\,fb$^{-1}$ accumulated
at the $\Upsilon(4S)$ resonance with the Belle detector at the KEKB
asymmetric-energy collider
corresponding to $\BelleNbb\times10^6$ $B\bar{B}$ pairs.

The Belle detector is a large-solid-angle magnetic
spectrometer that consists of a silicon vertex detector (SVD),
a 50-layer central drift chamber (CDC), an array of
aerogel threshold \v{C}erenkov counters (ACC),
a barrel-like arrangement of time-of-flight
scintillation counters (TOF), and an electromagnetic calorimeter (ECL)
comprised of CsI(Tl) crystals located inside
a superconducting solenoid coil that provides a 1.5~T
magnetic field.  An iron flux-return located outside of
the coil is instrumented to detect $K_L^0$ mesons and to identify
muons (KLM).  The Belle detector is described in detail elsewhere~\cite{belle}.
Two different inner detector configurations were used. For the first sample
of $152\times10^6$ $B\bar{B}$ pairs, a 2.0 cm radius beam-pipe
and a 3-layer silicon vertex detector were used;
for the latter $234\times10^6$ $B\bar{B}$ pairs,
a 1.5 cm radius beam-pipe, a 4-layer silicon detector
and a small-cell inner drift chamber were used~\cite{ushiroda}.
We use a GEANT based Monte Carlo (MC) simulation to model the
response of the detector and determine its acceptance\ \cite{sim}.

We detect the $\Lambda_c^+$ via the $\Lambda_c^+\to{pK^-\pi^+}$,
$p\bar{K^0}$ and $\Lambda\pi^+$ 
decay channels.
When a $\Lambda_c^+$ and ${\Lambda}_c^-$ are combined as $B$ decay
daughters, at least one of $\Lambda_c^{\pm}$ is required to have been
reconstructed via the $pK^{\mp}\pi^{\pm}$ decay process.
For each charged track, the particle identification (PID) information
from the CDC, ACC and 
TOF is used to construct likelihood functions $L_p$, $L_K$ and
$L_{\pi}$ for the proton, kaon and pion assignments, respectively.  
Likelihood ratios $L_a/(L_a+L_b)$ are required to be
greater than 0.6 to identify a particle 
as type $a$, where $b$ denotes the other two possible hadron assignments
from the three possiblities: proton, kaon and pion.
For the main mode $B^+\to\Lambda_c^+{\Lambda}_c^-K^+$,
$\Lambda_c^+\to{p}K^-\pi^+$, ${\Lambda}_c^-\to\bar{p}K^+\pi^-$
the PID efficiency for the primary $K^+$ is about $95\%$.
Efficiencies for protons, kaons and pions from $\Lambda_c^+$ decays
are about $98\%$.
The misidentification probability for pions (or kaons) to be identified
as kaons (or pions) is less than 5\%.
The probability for pions or kaons to be identified as protons is less
than 2\%.  
Tracks consistent with an electron or muon hypothesis are rejected.
A ${\Lambda_c^+}$ candidate is selected if the mass of its
decay products is within $0.010\GeV/c^2$ ($2.5\,\sigma$) 
of the nominal ${\Lambda_c^+}$ mass.

Neutral kaons are reconstructed in the $K_S^0{\to}\pi^+\pi^-$ decay.
Candidate $\Lambda$ baryons are reconstructed in
the decay $\Lambda{\to}p\pi^-$.
We apply vertex and mass constrained fits for the $K^0$ and $\Lambda$
candidates to improve the momentum resolution.
The intersection point of the $K^0$ and $\Lambda$
candidate daughter tracks must be displaced from the beam interaction
point (IP): the flight distance should be more than $0.5$\,mm.
A $K^0$ candidate is selected if the mass of its
decay products is within $7.5\MeV/c^2$ ($3\,\sigma$) 
of the $K^0$ mass.
A ${\Lambda}$ candidate is selected if the mass of its
decay products is within $2.5\MeV/c^2$ ($2.5\,\sigma$) 
of the ${\Lambda}$ mass.

The $B$ candidates are identified using the beam-energy
constrained mass $M_{\rm{bc}}$ and the mass difference
$\Delta M_B$.
The beam-energy constrained mass is defined as
$M_{\rm{bc}}\equiv\sqrt{E^2_{\rm{beam}}-(\sum\vec{p}_i)^2}$, where
$E_{\rm{beam}}$ is the beam energy, and $\vec{p}_i$ are the
three-momenta of the $B$ meson decay products, all defined in the
center-of-mass system (CMS) of the $e^+e^-$ collision. The mass
difference is defined as $\Delta{M}_B \equiv M(B)-m_{B}$,
where $M(B)$ is the reconstructed mass of the $B$ candidate
and $m_{B}$ is the world average $B$ meson mass. The parameter
$\Delta{M}_B$ is used instead of the energy difference
$\Delta{E}=(\sum E_i)-E_{\rm{beam}}$, where $E_i$ is the CMS energy of
the $B$ decay products, since $\Delta{E}$ shows a correlation with 
$M_{\rm{bc}}$, while $\Delta{M}_B$ does not~\cite{belle_zang}.
$M(B)=\sqrt{E(B)^2-(\sum\vec{p}_i)^2}$, where
$E(B)=E(\Lambda_c^+)+E({\Lambda}_c^-)+E(K)$, 
$E(\Lambda_c^+)=\sqrt{{\vec{p}_{\Lambda_c^+}}^{~2}+m_{\Lambda_c^+}^2}$,
$\vec{p}_{\Lambda_c^+}$ is the $\Lambda_c^+$ momentum measured via its
decay products and $m_{\Lambda_c^+}$ is 
the value of the $\Lambda^+_c$ baryon mass~\cite{lcmc_babar}.
We select events with $M_{\rm{bc}}>5.20$\GeV/$c^2$ and
$|\Delta{M}_B|<0.20$\GeV/$c^2$. 
The prompt $K^+$ or the reconstructed $K_S^0$ trajectory and the 
${\Lambda_c^+}/{\Lambda}_c^-$ trajectories are required to form a common
$B$ decay vertex. 
If there are multiple candidates in an event, the candidate with
the best $\chi^2_{B}$ for the $B$ vertex fit is selected.
The $B$ vertex fit is performed without additional
mass constraints for known particles.

\begin{figure*}[htb]
  \begin{tabular}{cc}
    \begin{minipage}{0.48\textwidth}
\psfrag{(b)}{\hspace{-0.25cm} (a)}
\psfrag{(c)}{\hspace{-0.25cm} (b)}
     \includegraphics[width=\textwidth]{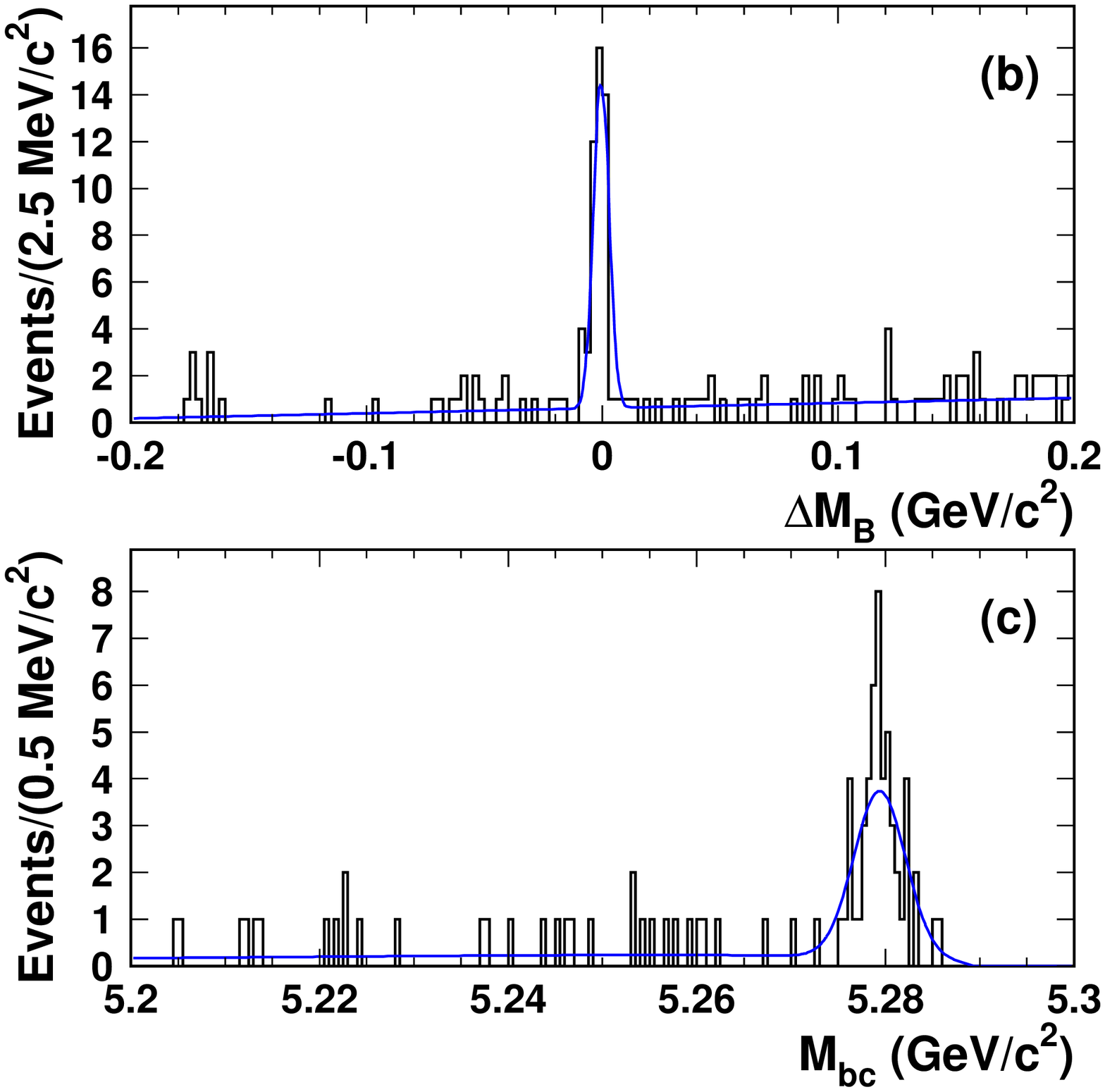}
    \end{minipage}
    &
    \begin{minipage}{0.48\textwidth}
\psfrag{(b)}{\hspace{-0.25cm} (c)}
\psfrag{(c)}{\hspace{-0.25cm} (d)}
      \includegraphics[width=\textwidth]{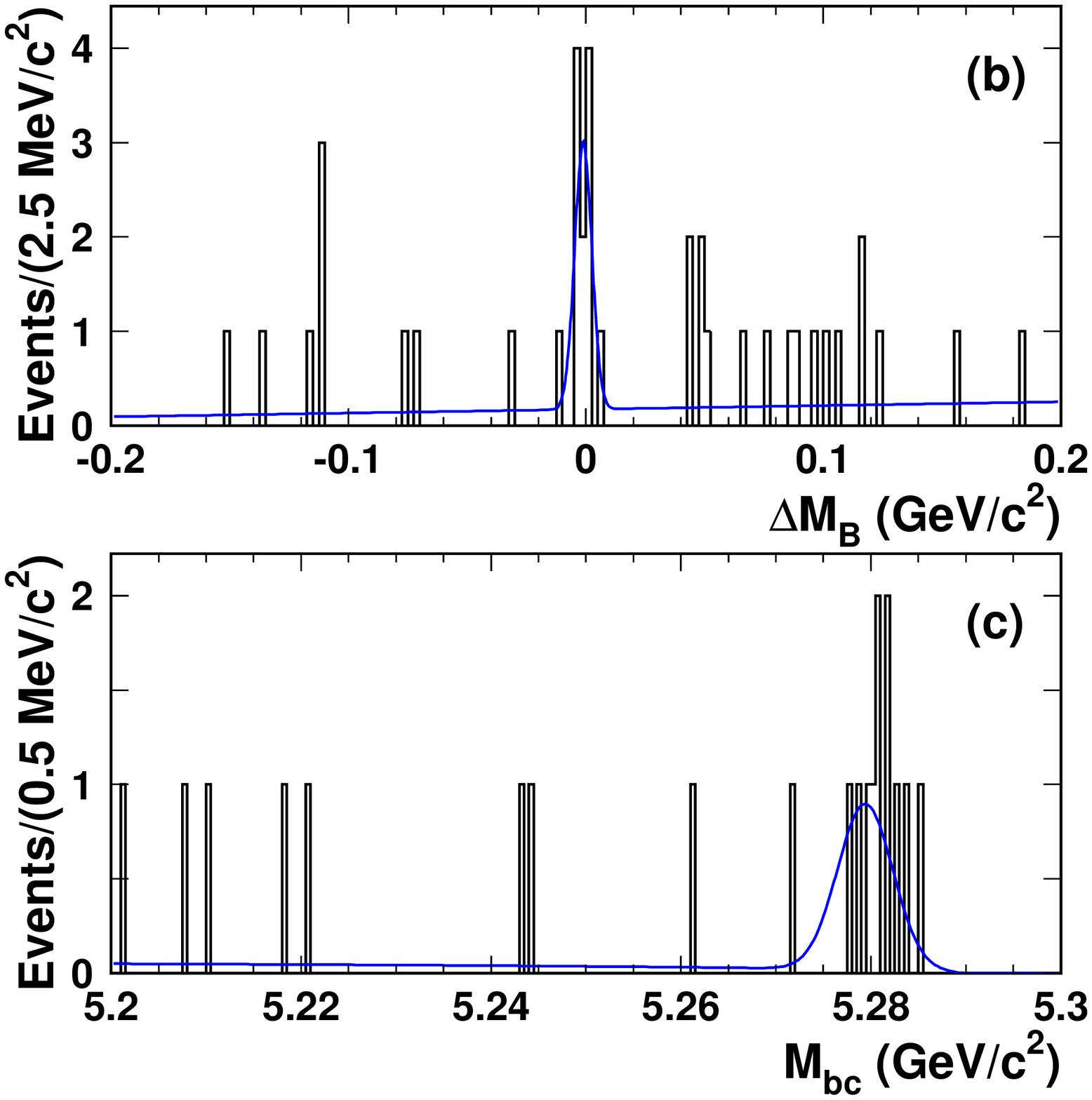}
    \end{minipage}
    \\
  \end{tabular}
  \caption{ Candidate  
    (a, b) ${{B^+}\to\Lambda_c^+{\Lambda}_c^-{K}^+}$ and 
    (c, d) ${{B^0}\to\Lambda_c^+{\Lambda}_c^-{K}^0}$ decay events: 
    (a, c) $\Delta{M}_B$ distribution for $M_{\rm{bc}}>5.27\GeV/c^2$ and
    (b, d) $M_{\rm{bc}}$ distribution for $|\Delta{M}_B|<0.015\GeV/c^2$.
    Curves indicate the fit results.
  }
  \label{fig:expsum}
\end{figure*}
Figure~\ref{fig:expsum} shows $\Delta{M}_B$ and $M_{\rm{bc}}$
projections for selected 
${{B^+}\to\Lambda_c^+{\Lambda}_c^-{K}^+}$ and
${{B^0}\to\Lambda_c^+{\Lambda}_c^-{K}^0}$ 
decay events.  
The $\Delta{M}_B$ projection is shown for $M_{\rm{bc}}>5.27\GeV/c^2$ and 
the $M_{\rm{bc}}$ projection is shown for $|\Delta{M}_B|<0.015\GeV/c^2$. 
The widths determined from single Gaussian fits to MC generated 
events are $2.7\,\MeV/c^2$ and $3.3\,\MeV/c^2$ for $M_{\rm{bc}}$
and $\Delta{M}_B$, respectively.
A two-dimensional binned maximum likelihood fit is performed to
determine the signal yield.
The $\Delta{M}_B$ distribution is approximated by 
a Gaussian for the signal plus a first order polynomial
for the background, and the $M_{\rm{bc}}$ distribution is represented 
by a single Gaussian for the signal plus an ARGUS
function\,\cite{argus_function} for the background. 
The signal shape parameters are fixed to the values
obtained from a fit to a MC simulation.
All yields and background shape parameters are allowed to float.

From the fit
we obtain signal yields of $\Nexplclckp$
and $\Nexplclcks$ events 
with statistical significances of $\Sexplclckp\sigma$ and $\Sexplclcks\sigma$, 
for ${{B^+}\to\Lambda_c^+{\Lambda}_c^-{K}^+}$ and
${{B^0}\to\Lambda_c^+{\Lambda}_c^-{K}^0}$, respectively. 
The significance is calculated as
$\sqrt{-2\ln({\cal{L}}_0/{\cal{L}}_\mathrm{max})}$, where
${\cal{L}}_\mathrm{max}$ 
and ${\cal{L}}_0$ denote the maximum likelihoods with the fitted signal
yield and with the yield fixed at zero, respectively.

The branching fraction ${\cal{B}}_{ij}$ for the $i$-th $\Lambda_c^+$
decay and the $j$-th ${\Lambda}_c^-$ decay
mode are calculated as
${\cal{B}}_{ij}=N_{ij}/[{N_{B\bar{B}}} \varepsilon_{ij}
 {\cal{B}}_i(\Lambda_c^+)  {\cal{B}}_j({\Lambda}_c^-)]$,
where $N_{ij}$ is the $B$ signal yield.
The detection efficiencies, $\varepsilon_{ij}$, are determined from MC
simulation.
The $\Lambda_c^+$ decay branching fractions ${\cal{B}}_i(\Lambda_c^+)$
are converted to the product
${\cal{B}}(\Lambda_c^+\to{pK^-\pi^+})
\Gamma_i/\Gamma({pK^-\pi^+})$  
to isolate the common uncertainty from the
branching fraction of $\Lambda_c^+\to{pK^-\pi^+}$.
The values of $\Gamma_i/\Gamma({pK^-\pi^+})$ are $(0.47\pm0.04)$ and
$(0.180\pm0.032)$ 
for $pK^0$ and $\Lambda \pi^+$ modes, respectively~\cite{pdg}.
The overall detection efficiency $\varepsilon$ for the total signal
yield $N$ is calculated as
$\sum{\varepsilon_{ij}
  [\Gamma_i/\Gamma({{p}K^-\pi^+})]
  [\Gamma_j/\Gamma({{p}K^-\pi^+})]}$.  
The overall branching fraction is calculated as $N_{\rm S}/[
 {N_{B\bar{B}}} \varepsilon {\cal{B}}(\Lambda_c^+\to{pK^-\pi^+})^2]$, 
using the overall signal yield $N_{\rm S}$ and
the decay branching fraction
${\cal{B}}(\Lambda_c^+\to{pK^-\pi^+})=(5.0\pm1.3)\%$~\cite{pdg}.
The detection efficiencies are calculated to be
$\EFFlclckp\%$ for the ${{B^+}\to\Lambda_c^+{\Lambda}_c^-{K}^+}$ decay
and 
$\EFFlclcks\%$ for the ${{B^0}\to\Lambda_c^+{\Lambda}_c^-{K}^0}$ decay.

The number of $B\bar{B}$ pairs $N_{B\bar{B}}$ is
$(\BelleNbb\pm\BelleNbbErr)\times{10^6}$.
The fractions of charged and neutral $B$ mesons are assumed to be equal. 
We obtain branching fractions of
\begin{center}
$
{\cal{B}}({{B^+}\to\Lambda_c^+{\Lambda}_c^-{K}^+})=
(\BFlclckp)\times10^{-4}~\mbox{and}
$\\
$
{\cal{B}}({{B^0}\to\Lambda_c^+{\Lambda}_c^-{K}^0})=
(\BFlclcks)\times10^{-4}, 
$
\end{center}
where the first and the second errors are statistical and systematic,
respectively.
The last error is due to the 52\% uncertainty in
the absolute branching fraction,
${\cal{B}}(\Lambda_c^+\to{p}K^-\pi^+)$. 

Systematic uncertainties in the detection efficiencies arise from
the track reconstruction efficiency (8\% -- 10\% depending on
the process, assuming a correlated systematic error of about 1\% per
charged track);
the PID efficiency (9\% -- 10\% assuming a correlated systematic error of
2\% per proton and 1\% per pion or kaon);
3-body decay model uncertainty (11\% for
the ${{B^+}\to\Lambda_c^+{\Lambda}_c^-{K}^+}$ decay and
5\% for the ${{B^0}\to\Lambda_c^+{\Lambda}_c^-{K}^0}$ decay);
and MC statistics (1\% -- 2\%).
The other uncertainties are associated with
$\Gamma(\Lambda_c^+)/\Gamma({pK^-\pi^+})$
(2\% -- 3\%); and
the number of ${N_{B\bar{B}}}$ events (1\%).
The total systematic error is 17\% for 
$B^+\to\Lambda_c^+\Lambda_c^- K^+$  and 15\% for
$B^+\to\Lambda_c^+\Lambda_c^- K^0$.

\begin{figure*}[htb]
  \begin{tabular}{cc}    
    \begin{minipage}{0.48\textwidth}
\psfrag{(a)}{\hspace{-0.25cm} (a)}
      \includegraphics[width=\textwidth,bb=0 0 567 296]{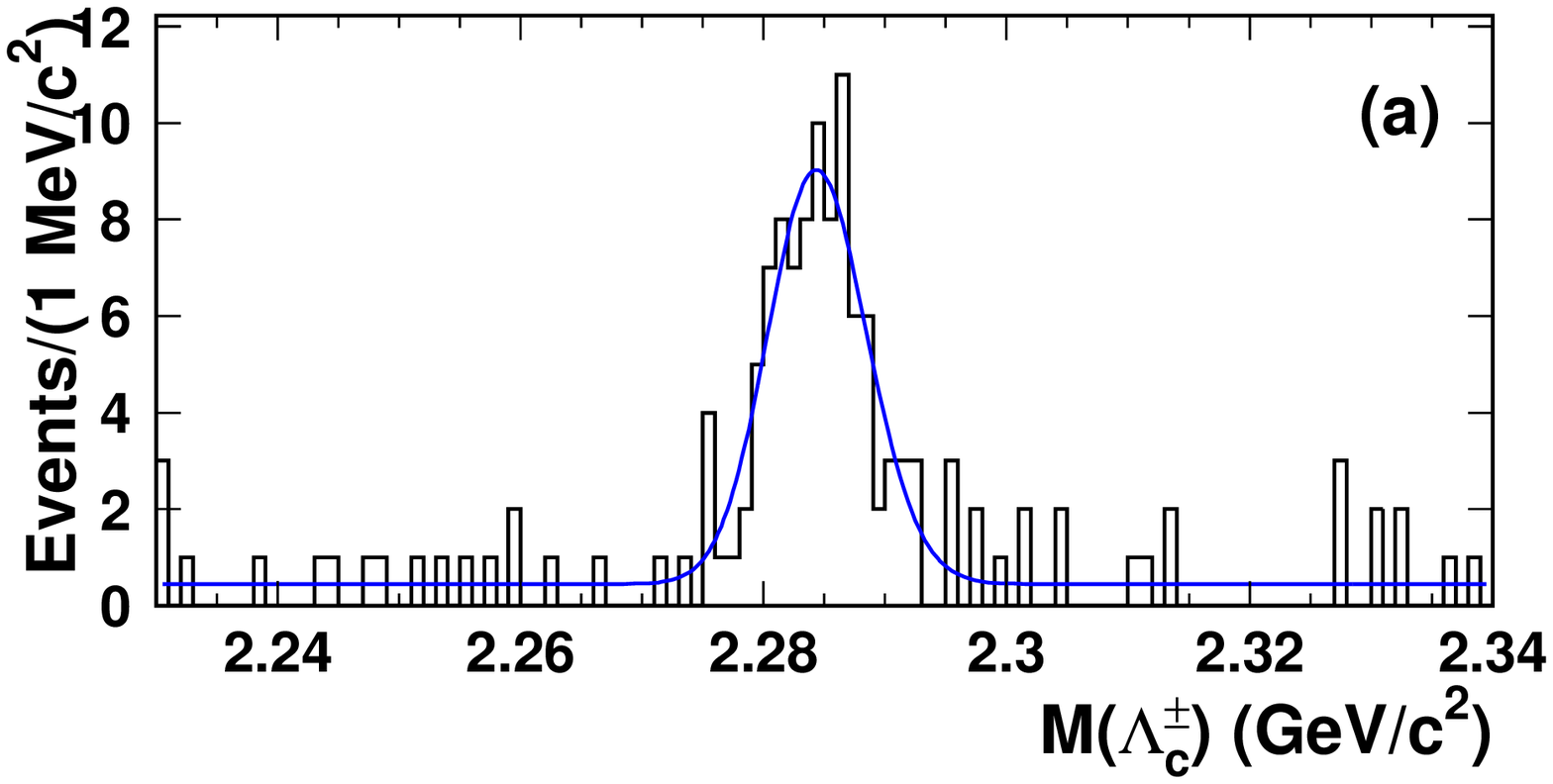}
    \end{minipage}
    &
    \begin{minipage}{0.48\textwidth}
\psfrag{(b)}{\hspace{-0.25cm} (b)}
      \includegraphics[width=\textwidth,bb=0 0 567 296]{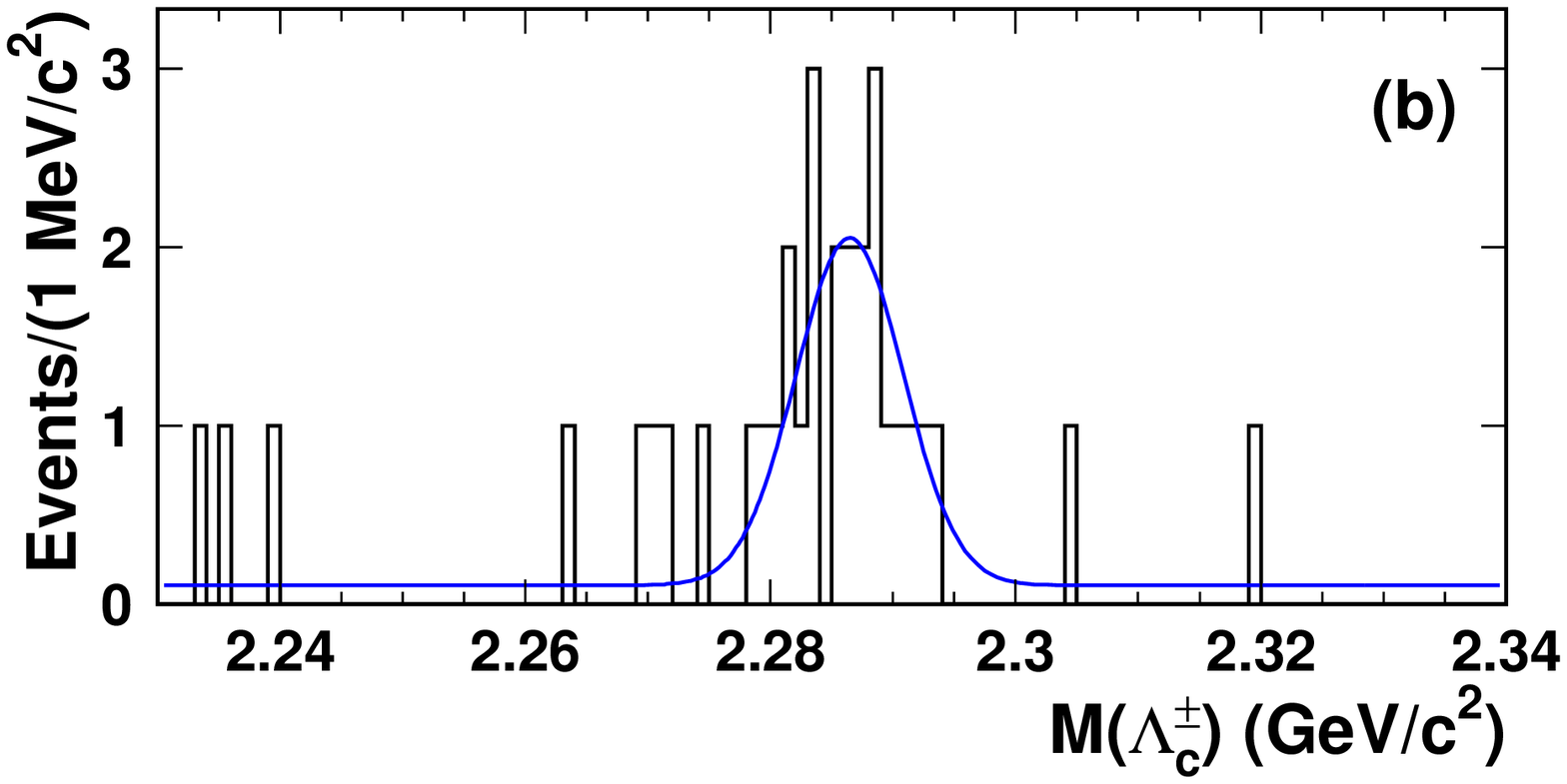}
    \end{minipage}
    \\
  \end{tabular}
  \caption{
    $M(\Lambda_c^{\pm})$ mass distributions for 
    (a) ${{B^+}\to\Lambda_c^+{\Lambda}_c^-{K}^+}$ and
    (b) ${{B^0}\to\Lambda_c^+{\Lambda}_c^-{K}^0}$ decay candidates
    in the $B$ signal region. 
    Curves indicate the fit results.
  }
  \label{fig:lcmass}
\end{figure*}

Figure~\ref{fig:lcmass} shows the mass distributions
$M(\Lambda_c^{\pm})$ for $B$ candidates in the signal region
$|\Delta{M}_B|<0.015\GeV/c^2$ and $M_{\rm{bc}}>5.27$\GeV/$c^2$. 
The $M(\Lambda_c^{\pm})$ mass distributions are shown for
$|M({\Lambda}_c^{\mp})-m_{\Lambda_c^+}|<0.010\GeV/c^2$.
The curves show the results of a fit with the sum of a Gaussian
and a linear background. The means and widths of the Gaussians
are fixed to values obtained from fits to MC samples.
For ${{B^+}\to\Lambda_c^+{\Lambda}_c^-{K}^+}$ decay,
we obtain a ${\Lambda_c^+}$ yield of $\NexplclckpLcp$ events and
a ${{\Lambda}_c^-}$ yield of $\NexplclckpLcm$ events.
For ${{B^0}\to\Lambda_c^+{\Lambda}_c^-{K}^0}$, yields of 
$\NexplclcksLcp$ and 
$\NexplclcksLcm$ events are obtained from
the ${\Lambda_c^+}$ and ${{\Lambda}_c^-}$ distributions, respectively.
These values are consistent with the $B$ signal yields given above.


We consider possible contributions from other $B$ decays, 
which could give a $B$ signal in the $\Delta{E}$ and $\Delta{M}_B$
distributions,  
but should produce a uniform distribution in the $\Lambda_c^+$ 
mass region.  
To assess this type of background, we analyze the $\Lambda_c^+$ sideband 
$0.015\GeV/c^2<|M(\Lambda_c^+)-m_{\Lambda_c^+}|<0.055\GeV/c^2$ and
$|M({\Lambda}_c^-)-m_{\Lambda_c^-}|<0.010\GeV/c^2$,
and ${\Lambda}_c^-$ sideband 
$0.015\GeV/c^2<|M({\Lambda}_c^-)-m_{\Lambda_c^-}|<0.055\GeV/c^2$ and
$|M({\Lambda}_c^+)-m_{\Lambda_c^+}|<0.010\GeV/c^2$.
We conclude that other $B$ decays contribute less than
$\ULexplclckpLcsideLcsign$ events 
at 90\% C.L. in the ${{B^+}\to\Lambda_c^+{\Lambda}_c^-{K}^+}$  mode
and less than $\ULexplclcksLcsideLcsign$ events at 90\% C.L. in
${{B^0}\to\Lambda_c^+{\Lambda}_c^-{K}^0}$; both contributions are
neglected.

\begin{figure*}[htb]
  \begin{tabular}{cc}
    \begin{minipage}{0.48\textwidth}   
\psfrag{(a)}{\hspace{-0.25cm} (a)}
      \includegraphics[width=\textwidth,bb=0 0 567 296]{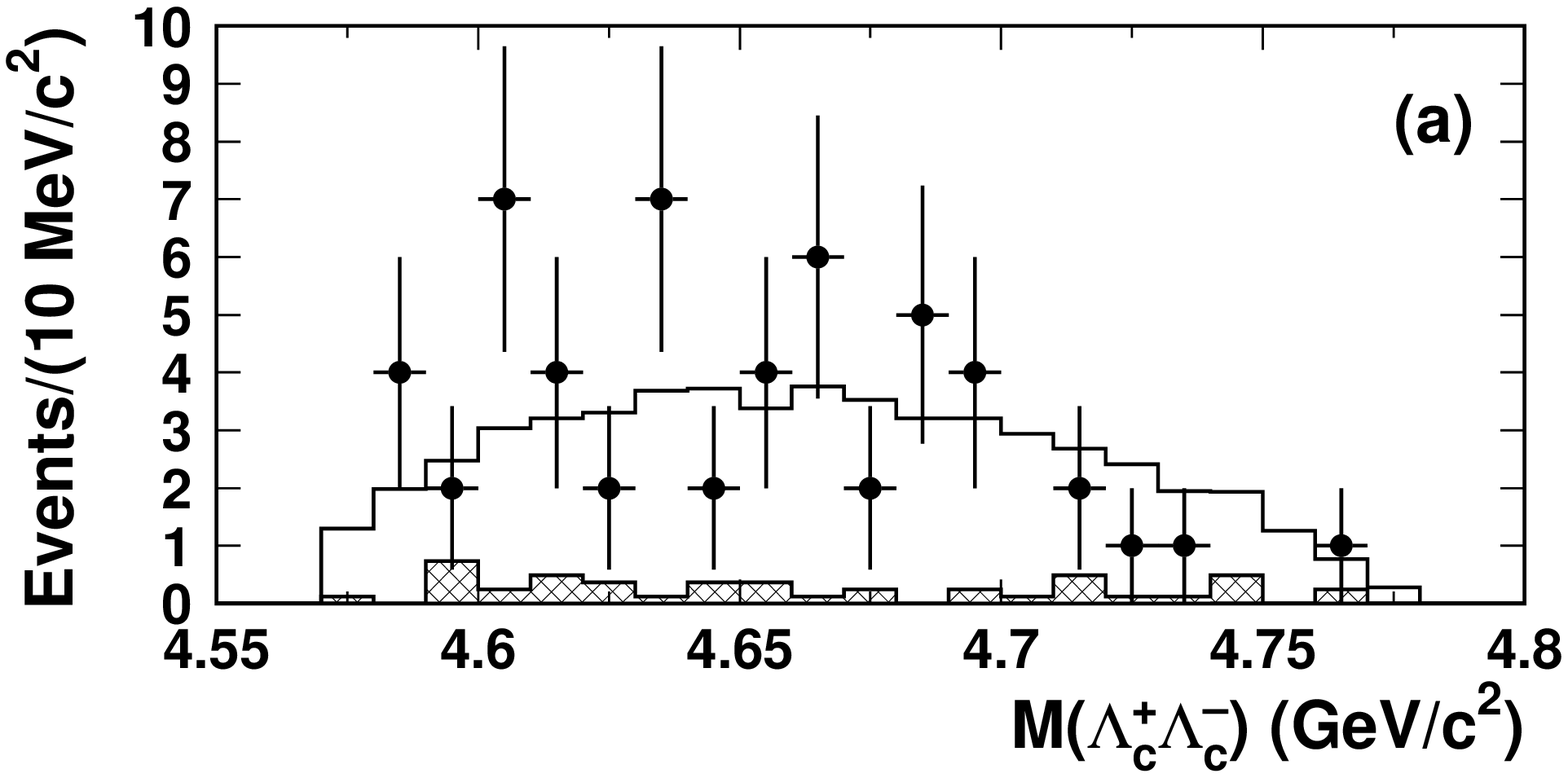}
    \end{minipage}
    &
    \begin{minipage}{0.48\textwidth}   
\psfrag{(b)}{\hspace{-0.25cm} (b)}
      \includegraphics[width=\textwidth,bb=0 0 567 296]{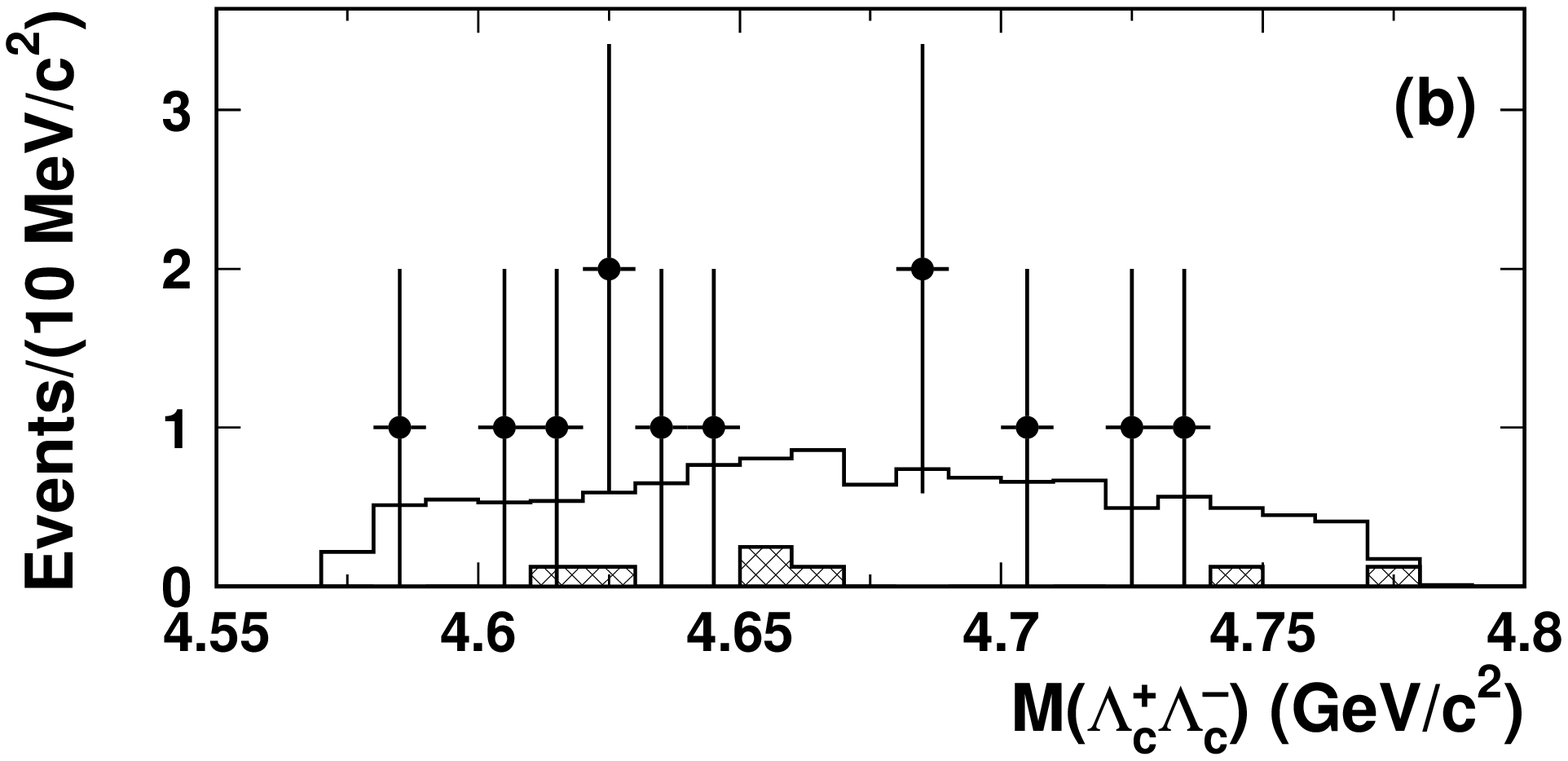}
    \end{minipage}
    \\
  \end{tabular}
  \caption{ 
$M({\Lambda}_c^+{\Lambda}_c^-)$ mass distributions for
(a) ${{B^+}\to\Lambda_c^+{\Lambda}_c^-{K}^+}$ and 
(b) ${{B^0}\to\Lambda_c^+{\Lambda}_c^-{K}^0}$ decay candidates 
in the $B$ signal region. 
Points with error bars are data.
Open histograms -- MC for a uniform phase space distribution.
Hatched histograms -- normalized $\Lambda_c^+$ sideband data.
  }
  \label{fig:minv_lclc}
    \end{figure*}
Figure~\ref{fig:minv_lclc} shows 
the $M({\Lambda}_c^+{\Lambda}_c^-)$ mass distributions for
(a) ${{B^+}\to\Lambda_c^+{\Lambda}_c^-{K}^+}$ decay candidates and 
(b) ${{B^0}\to\Lambda_c^+{\Lambda}_c^-{K}^0}$ decay candidates 
in the $B$ signal region $|\Delta{M}_B|<0.015\GeV/c^2$ and
$M_{\rm{bc}}>5.27$\GeV/$c^2$. 
No deviations from phase space distributions are evident.

In summary,
we have reported the first measurement of the doubly charmed baryonic $B$ decay
${{B^+}\to\Lambda_c^+{\Lambda}_c^-{K}^+}$ with 
a branching fraction of 
$(\BFlclckp)\times10^{-4}$
and a statistical significance of $\Sexplclckp\,\sigma$, and
the ${{B^0}\to\Lambda_c^+{\Lambda}_c^-{K}^0}$ decay with
a branching fraction of 
$(\BFlclcks)\times10^{-4}$
and a statistical significance of $\Sexplclcks\,\sigma$.
These three-body doubly charmed $B$ decay branching fractions are 
about the same order of magnitude (or slightly smaller)
than the branching
fraction of the two-body doubly charmed decay
$B^+\to\bar{\Xi}_c^0\Lambda_c^+$, which is due to the same 
$b{\to}c\bar{c}s$ quark transition, also observed by
Belle~\cite{belle_xic_lamc}. 
The behavior of these $b{\to}c\bar{c}s$ decays is qualitatively
different from single charmed baryon decays, 
where three-body decays have bigger branching fractions 
than two-body decays.
The obtained branching fraction is by five to six orders of magnitude
higher than expected from naive estimation for the
${{B}\to\Lambda_c^+{\Lambda}_c^-{K}}$ decay with color suppression, 
which is also highly suppressed by phase space~\cite{puzzle}.
All this needs further experimental and theoretical study.

\begin{acknowledgments}
We thank the KEKB group for excellent operation of the
accelerator, the KEK cryogenics group for efficient solenoid
operations, and the KEK computer group and
the NII for valuable computing and Super-SINET network
support.  We acknowledge support from MEXT and JSPS (Japan);
ARC and DEST (Australia); NSFC and KIP of CAS (contract No.~10575109
and IHEP-U-503, China); DST (India); the BK21 program of MOEHRD, and
the
CHEP SRC and BR (grant No. R01-2005-000-10089-0) programs of
KOSEF (Korea); KBN (contract No.~2P03B 01324, Poland); MIST
(Russia); MHEST (Slovenia);  SNSF (Switzerland); NSC and MOE
(Taiwan); and DOE (USA).
\end{acknowledgments}

\end{document}